\documentclass[11pt]{article}
\usepackage{graphicx}
\usepackage{placeins}
\usepackage{float}
\usepackage{subfigure}
\usepackage[toc,page]{appendix}

\usepackage{amssymb}
\usepackage{epsfig}
\begin{document}
\title{{\bf{\Large Hawking evaporation cascade in presence of back reaction effect}}}
\author{ 
{\bf {\normalsize Avik Paul}$
$\thanks{E-mail: avik.paul@iitg.ernet.in}} \,\ and \,\
 {\bf {\normalsize Bibhas Ranjan Majhi}$
$\thanks{E-mail: bibhas.majhi@iitg.ernet.in}}\\ 
{\normalsize Department of Physics, Indian Institute of Technology Guwahati,}
\\{\normalsize Guwahati 781039, Assam, India}
\\[0.3cm]
}

\maketitle

\begin{abstract}
We study the cascade of Hawking emission spectrum from the event horizon in presence of one loop back reaction effect in a black hole background. The spacetime is taken here is the modified Schwarzschild one. The analysis shows that it is possible to decrease the sparsity with the decrease in black hole mass. Moreover, at some particular value of mass one has a continuous radiation cascade. This result is completely new and quite different from the usual one. An estimation of the mass for continuous one is also found. We see that the value is of the Planck mass order. In this process it is observed that under a physical background, below a particular value of the mass the Hawking radiation must stop and we have a remnant. This was absent in the earlier analysis.
\end{abstract}

\section{Introduction}
    Semi-classical treatment shows that black holes radiate and the emission spectrum, if one ignores the gray body factor, is analogous to thermal black body radiation \cite{Hawking:1974sw}. Just after this discovery, Page's calculation \cite{Page:1976df,Page:1976ki} revealed that the time gap between the successive emission is very large compared to a characteristic time scale of the system. Very recently, the same issue has been studied in a thermodynamical approach by Gray et. al. in \cite{Gray:2015pma} and the conclusion is again identical -- {\it the cascade of Hawking emission is very sparse}. They showed that the average time gap between the emission of successive particle is remain constant with respect to the natural time scale and the value is quite large compared to unity. 
In this recent analysis the authors took the usual black holes as background so that the temperature and area were given by standard values. 

  In this paper, adopting the identical formalism we will study the same issue when the background black hole metric is modified by the back reaction. Here we shall mainly consider the modified Schwarzschild black hole spacetime, obtained in \cite{Lousto:1988sp}, due to one loop calculation. It has been observed that due this effect the temperature of the horizon and area changes and are corrected by the term which are of the $\hbar$ order. Since in principle the fluctuations of the external fields and $g_{ab}$, known as back reaction, affects the underlying backgounds, it would be interesting to see the nature of the Hawking emission cascade in presence of such effects. We shall precisely address this issue here. 

  We shall show that it is possible to obtain continuous cascade of emission for a particular value of mass of the black hole. Above this, the cascade is becoming sparse and the sparsity increases with the increase of mass. In addition to that it will be observed that at a particular value of mass, the ratio between the successive emission time and the characteristic time scale becomes zero and below this it is negative. Since such thing is physically impossible, one can conclude that the black hole will stop evaporating at this precise value of mass and there will not be a complete evaporation; i.e. one will have a Planck size remnant. All these features are completely new and very interesting as the Planck scale physics is very important in context of gravity.

   The organization of the paper is as follows. In the next section, a summary of the results from \cite{Gray:2015pma} will be presented which will be necessary for our main purpose. Section \ref{back} will contain our main analysis and results. Finally we shall conclude.

\section{\label{review}A brief review: General framework}
  In this section, for our main purpose we summarise the required basic results, discussed very recently in \cite{Gray:2015pma}. These could be borrowed directly from this paper, but for the shake of completeness and transparency  a brief discussion will be presented here. We shall note that, such a discussion helps us to understand the generality and applicability of these results for different cases. In this way, the feasibility of applying the results directly in our present paper will be justified.  In addition to that the same will be useful for a new reader. 

  The Hawking radiation from the black hole horizon is pure thermal radiation if one neglects the grey body factor \cite{Hawking:1974sw}.  Depending upon the nature of the particle (Bosons or Fermions), the distribution is either Bose-Einstein or Fermi-Dirac. The number distribution, obeying the Bose-Einstein statistics, is given by
  \begin{equation}
  \bar{n}= \frac{1}{\exp\Big(\frac{\hbar k c}{k_{B} T}\Big)-1}~,
  \label{n}
  \end{equation}
  where $\hbar$, $k_B$ and $c$ are Planck's constant, Boltzmann constant and velocity of light, respectively; while $k$ and $T$ are wave number and Temperature of the horizon, respectively.
 So total number of bosons emitted through area $A$ per unit time; i.e. the rate of emission is 
\begin{eqnarray}
\Gamma = \frac{g}{2\pi^2}\int_{0}^{\infty} \bar{n}c k^2A dk = \frac{gk_{B}^3 T^3}{\pi^2\hbar^3 c^2}\zeta(3)A~. 
\label{4} 
\end{eqnarray}
(Detailed calculation can be followed from \cite{Pathria}. See Eq. (23) of page 172). Therefore, time taken for the emission of one boson is the inverse of the above quantity.  Hence this can be considered as the average time gap between the emission of successive photons which we denote as $\tau_{gap} = \frac{1}{\Gamma}$.  Next we want to compare such time with the characteristic time scale of the emission process.  

      The characteristic time scale of the process can be determined by the frequency at which the integrand of (\ref{4}) is maximum; i.e. the number of emitted bosons is maximum. One can easily verify that the corresponding value of wave number is given by \cite{Gray:2015pma}
\begin{equation}
k_{peak}= \frac{k_{B}T}{\hbar c}\Big(2+W(-2e^{-2})\Big)~, 
\label{5}
\end{equation}
where $W(x)$ is the Lambart $W$- function which is defined as $W(x)e^{W(x)} = x$.  Therefore, the frequency at which the distribution of photon number has peak turns out to be
\begin{equation}
\nu_{peak}=\frac{\omega_{peak}}{2\pi} = \frac{ck_{peak}}{2\pi}=\frac{k_{B}T}{2\pi \hbar }\Big(2+W(-2e^{-2})\Big)~.
\label{nu}
\end{equation} 
Hence the characteristic time scale of the system is set to be $\tau_{localization} = 1/\nu_{peak}$.  Now it is time to calculate $\tau_{gap}$ and $\tau_{localization}$ by using (\ref{4}) and (\ref{nu}) respectively, for a black hole emission process and measure $\tau_{gap}$ in units of $\tau_{localization}$. This will give the characteristic nature of the cascade of Hawking process. For that it is useful to define a dimensionless  quantity $\eta$ as
\begin{equation}
\eta =\frac{\tau_{gap}}{\tau_{localization}} = \frac{\nu_{peak}}{\Gamma}~,
\label{eta}
\end{equation}
which with the use of (\ref{4}) and (\ref{5}), turns out to be 
  \begin{equation}
  \eta = \frac{\pi\hbar^2c^2}{2g\zeta(3)k_{B}^2T^2A}(2+W(-2e^{-2}))~.  
  \label{6}
  \end{equation}
Let us now finish this with the following comments which will be useful for our later purpose. Because $\tau_{gap}$ is the average time gap between two successive emission of boson and $\tau_{localization}$ is the characteristic time scale of emission of individual particle, we always have a bound on the value of $\eta$, which is $\eta$ must be positive; i.e. 
\begin{equation}
\eta > 0~.
\label{main}
\end{equation}  
Such a bound coming from the fact that in a system the time taken for a process and the characteristic time scale, both must be positive.  When a process is continuous we must have $\eta\simeq\mathcal{O}(1)$ while one calls this process as sparse when $\eta > \mathcal{O}(1)$. Moreover, $0<\eta<\mathcal{O}(1)$ is possible if the surface area of the emitter is very very large compare to the wavelength of the emitted particle.
Also remember that the expression (\ref{6})  is very much general for any black holes in ($1+3$) dimensions  and one considers the pure thermal radiation as bosons from the horizon. Therefore, we take this one as our working expression for our main discussion.

    For schwarzschild black hole $k_{B}T = \hbar c/4\pi r_{H}$  and $A=A_H=4 \pi r_{H}^2$ and hence
 \begin{eqnarray}
\eta =\frac{2\pi^2(2+W(-2e^{-2}))}{g\zeta(3)} = \frac{26.182}{g}~,   
\label{7}
 \end{eqnarray}
where $W(-2e^{-2})= -0.406$ and $\zeta(3)= 1.202$ have been substituted.
Note that the above result is independent of black hole parameter; e.g. mass here. Moreover, it is just a pure number. Also, observe that the value of $\eta$ is not comparable to order unity; rather that is far from unity; i.e. $\eta>>>\mathcal{O}(1)$.  Therefore one can conclude that the usual Hawking radiation in this case is very sparse. This result is not new; the similar nature has also been obtained recently in \cite{Gray:2015pma}. The only difference is the numerical value. The reason for that is as follows.
If one incorporates the grey body factor in (\ref{n}), then the correct value of effective area of emission in the high frequency limit is $A=(27\pi r_H^2)/4 = (27A_H)/16$ \cite{Page:1976df}. Use of this in (\ref{6}) leads to the result that obtained in \cite{Gray:2015pma}. This has precisely been done by Gray et. al. It can be observed that with this corrected value, the above conclusion does not change. So it is sufficient not to take into account the grey body factor in the analysis in order to obtain the characteristic nature of the value of $\eta$.  Hence in the later analysis, with out loss of generality, the area of emission $A$ for a spherically symmetric black hole will be taken to be $A=4\pi r_H^2$ where $r_H$ will be identified as the horizon radius.

\section{\label{back}Effect of back reaction} 
   Due to the fluctuation of external fields including graviton on a classical curved background, a non-zero expectation value of energy-momentum tensor, at the one loop level, appears. If one includes this on the right hand side of Einstein's equation, the usual solution gets modified. This is known as back reaction problem. It has been observed that under such circumstances, the usual form of horizon temperature and entropy of a black hole solution of the {\it semi-classical} Einstein's equation include a correction at the Planck's constant $\hbar$ order level \cite{Lousto:1988sp}, \cite{Fursaev:1994te}--\cite{Banerjee:2008gc}. Such corrections usually regarded as one type of quantum correction. 
   
   For the Schwarzschild black hole case, the modified temperature, upto one loop calculation, is given by \cite{Lousto:1988sp}, \cite{Banerjee:2008ry}--\cite{Majhi:2009uk},
\begin{equation}
T= \frac{\hbar}{4\pi r_{H}} \Big(1+ \frac{\alpha \hbar}{r_H^2} \Big)~. 
\label{9}
\end{equation}
where $\alpha$ is dimensionless parameter and related to the trace anomaly coefficient. In the above we have chosen our unit such that $c=1=k_B$. The explicit value of $\alpha$ has been derived in literature \cite{Page:2004xp}: 
\begin{equation}
\alpha = \frac{1}{360 \pi}\Big(-N_{0}-\frac{7}{4}N_{\frac{1}{2}} +13N_{1}+ \frac{233}{4}N_{\frac{3}{2}} -212N_{2}\Big)
\label{alpha}
\end{equation}
where $N_{s}$ denotes the number of fields with spin $s$. It is has already been shown that such quantum correction leads to modifications to the horizon entropy. The leading order correction is logarithmic of usual area whose coefficient is given the trace anomaly (\ref{alpha}). For details, see Refs. \cite{Fursaev:1994te}--\cite{Majhi:2009uk}.   
After specifying the modified temperature, it is necessary to mention the correction to horizon radius to find the horizon area which will be substituted in Eq. (\ref{6}) to obtain the modification to $\eta$. The corrected value of the horizon radius at the one loop order turned out to be \cite{Lousto:1988sp}
\begin{equation}
r_h= r_{H} \Big(1+ \frac{\beta \hbar}{r_H^2}\Big) 
\label{10}
\end{equation}
with $\beta$ is also dimensionless parameter. The explicit value of $\beta$ depends on the time-time component of regularised energy-momentum tensor (See Eq. (23) of \cite{Lousto:1988sp}).

    Now it is time to calculate $\eta$ for this situation. To use Eq. (\ref{6}) one has to make sure that Eq. (\ref{n}) is still valid in presence of back reaction effect. In this regard, we want to point out that the quantum mechnically corrected Schwarzschild black hole is spherically symmetric and static, like the usual one (see Ref. \cite{Lousto:1988sp}, for details). The only modification is in the metric coefficients. Therefore, it is obvious that the usual form of the result for the emission spectrum will also be obtained here. The modifications are reflected in the precise expression for temperature.
In order to calculate $\eta$ by using Eq. (\ref{6}) where one has to substitute (\ref{9}) and $A=4\pi r_h^2$. Since the results (\ref{9}) and (\ref{10}) are valid up to oder $\hbar$, we determine $\eta$ by retaining terms which are of the same order. This leads to
\begin{equation}
\eta = \frac{26.182}{g}\Big(1- \frac{2\hbar \gamma}{r_H^2} \Big)~,
\label{eta1}
\end{equation}
where the dimensionless constant $\gamma$ is defined as $\gamma = \alpha + \beta$. 
Note that $\eta$ is no longer a pure number, rather it depends on the black hole mass $M$ as $r_H=2M$. Hence the value of $\eta$ can vary with mass. Of course the nature of variation must depends on the value of dimensionless constant $\gamma$. This gives rise to three possibilities depending upon the values of $\eta$. 
\vskip 2mm
\noindent
{\underline{Case I}}: When $\gamma = 0$; i.e. $\alpha=-\beta$. \\
Then we find the same result ($\eta = 26.182/g$) like what was obtained in absence of back reaction (see Eq. (\ref{7})). So $\eta$ remains constant and as earlier, the Hawking cascade, in this case, is again very sparse.
\vskip 2mm
\noindent     
{\underline{Case II}}: For $\gamma > 0$.\\
Here we rewrite (\ref{eta1}) as 
\begin{equation}
\eta = \frac{26.182}{g}\Big(1- \frac{2\hbar |\gamma |}{r_H^2} \Big)~.
\label{eta2} 
\end{equation}
Note that this approaches to the usual constant value in the limit $r_H\rightarrow\infty$. But when $r_H$ decreases, the value of $\eta$ also decreases. Now note that the Hawking cascade will be continuous for the following value of $r_H$
\begin{equation}
\frac{r_H}{\sqrt{\hbar|\gamma|}}\simeq\mathcal{O}\Big(\Big[\frac{2}{1-\frac{g}{26.182}}\Big]^{1/2}\Big)~.
\label{minr} 
\end{equation}
This feature is completely new as in the usual analysis, where there is no back reaction, does not show such feature. It implies that in Hawking radiation a continuous cascade is possible when the classical radius of event horizon is of the order given by the right hand side of the above equation.
So the story is like this. Initially the Hawking radiation is not continuous. But since during evaporation the radius decreases, the radiation time between two particles is becoming more and more less. When $r_H$; i.e. the black hole mass reaches to the order of magnitude as calculated from the right hand side of the above equation, the nature of radiation is totally continuous. So we conclude that {\it in presence of back reaction effect a continuous cascade in Hawking emission is possible}.       
\vskip 2mm
\noindent    
{\underline{Case III}}: For $\gamma < 0$.\\
In this case one has the following expression:
\begin{equation}  
\eta = \frac{26.182}{g}\Big(1+ \frac{2\hbar |\gamma |}{r_H^2} \Big)~.
\label{eta3} 
\end{equation}
Here $\eta$ increases from its saturation value when $r_H$ decreases. So in this case the Hawking radiation is always sparse. For low values of $r_H$ it will be more sparse than the usual one.

     Now we want to give a graphical analysis for these three cases. In addition to that an estimation of $r_H$ for the continuous cascade corresponding to Case II can also be given. For that we need to know the value of degeneracy parameter $g$. Below let us consider that the emitted particles are photons. Since it has two modes of polarisation in this case $g=2$. Then the values of $\eta$ for the three possibilities will be as follows:
\begin{eqnarray}     
{\textrm{Case I}:} &&\eta = \frac{26.182}{2} = 13.091~;
\nonumber
\\
{\textrm{Case II}:} &&\eta = 13.091\Big(1- \frac{2}{r_H^2} \Big)~;
\nonumber
\\
{\textrm{Case III}:} &&\eta = 13.091\Big(1+ \frac{2}{r_H^2} \Big)~;
\label{all}
\end{eqnarray}
where in the above $r_H$ is taken in the unit of $\sqrt{\hbar |\gamma|}$; i.e $r_H/(\sqrt{\hbar|\gamma|})$ has been replaced by $r_H$. Below we give the  $\eta - r_{H}$ plot for three cases.
\begin{figure}[H]
  \centering
    \includegraphics[width=0.5\textwidth]{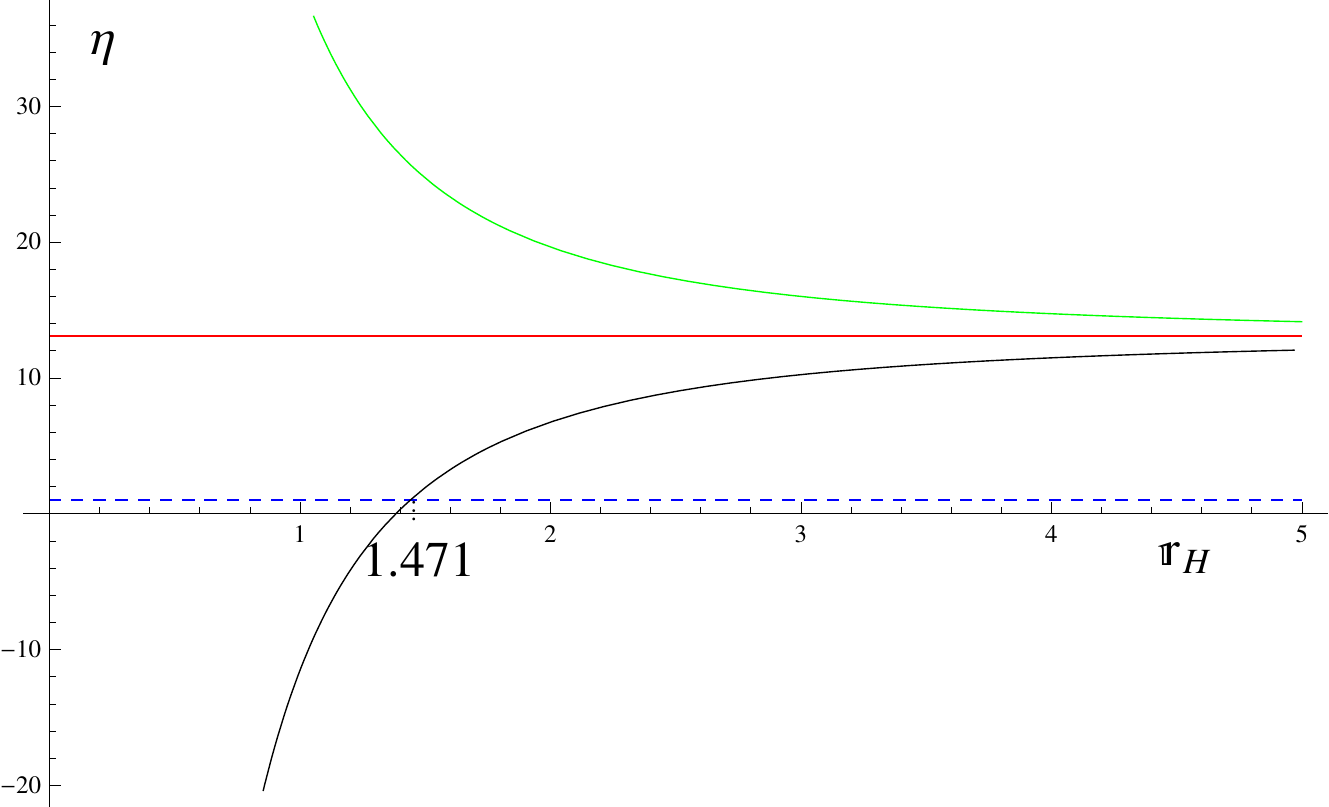}
     \caption{{\it{$\eta$ Vs $r_H$ plot.~~
Red line: Case I; Black line: Case II; Green line: Case III.}}}
\end{figure}
\noindent
Case I gives a straight line parallel to $r_H$ axis and it is far from $\mathcal{O}(1)$. In Case III, there is no upper bound on $\eta$. At lower value of $r_H$, $\eta$ increases rapidly while at higher value it becomes saturate to straight line obtained in Case I. Case II is the most interesting. At higher values of black hole mass it saturates to straight line, obtained in Case I; on the other hand at some lower value of $r_H$ , $\eta$ decreases. It is of the order unity when $r_{H_{(continuous)}} \simeq \mathcal{O}(1.471\sqrt{\hbar|\gamma|})$. So during the evaporation, when the mass of the black hole reaches $M_{continuous}\simeq\mathcal{O}(0.736\sqrt{\hbar|\gamma|})$; i.e. of the Planck mass order, we can see that the cascade of radiation is continuous with respect to the characteristic time scale given by (\ref{nu}). At other values which are above this point it is sparse and this sparsity increases with the increase of mass. 
Below this point we have much more dense cascade of radiation as the thermal wavelength $\lambda=(2\pi\hbar)/T$ becomes in the regime of Planck length order and so it is very very small compared to the emitter's surface area. This continues upto the limit $r_H\rightarrow1.414\sqrt{\hbar|\gamma|}$.

    Let us conclude case II by discussing one more important and new feature. From the temperature expression in presence of back reaction with $\alpha$ is positive, it has already been observed that for low value of $r_H$ the temperature diverges (see Eq. (\ref{9}) and for plot of it look at the refs. \cite{Lousto:1988sp,Banerjee:2008gc});  So it was concluded that when the black hole evaporates there will not be any remnant and it will go to singularity. In contrary to this earlier one, the present analysis shows it is indeed not the case. Note from the figure 1 that for Case II, $\eta$ becomes negative for $r_H<1.414\sqrt{\hbar|\gamma|}$.  But we already argued that $\eta$ can not be negative (see discussion around Eq. (\ref{main})). Therefore it might be possible that $\eta$ can be a candidate for determining the remnant of the evaporation of black hole like the temperature as prescribed in \cite{Lousto:1988sp}. So the black hole will must stop evaporating when $r_H$ reaches this value.  Here our underlying condition is $\gamma$ has to be positive and such phenomenon is possible even if $\alpha$ is positive. Hence we must have some remnant and it will not reach to singularity. Such conclusion is completely opposite what we learned from the earlier analysis which depends on the temperature expression. To achieve this different conclusion, we found that one has to impose a fundamental condition on $\eta$ given by Eq. (\ref{main}) and this is quite reasonable to consider as physically both the time scale can not be negative. Let us also mention that there are other ways to find the remnant for a black hole evaporation case; like the vanishing of the variation of the energy emitted per unit time; i.e. $dM/dt=0$, vanishing of the specific heat \cite{Banerjee:2010sd}, vanishing of the rate of change of horizon entropy with temperature \cite{Banerjee:2010sd} and the temperature of horizon must be real \cite{new2,Nozari:2005ah}. In some cases it may possible that the values of the remnant mass agrees in all methods while it is not guaranteed that it must hold in all times \cite{Banerjee:2010sd,new2,Nozari:2005ah,Nouicer:2007jg}. For example, in \cite{Nouicer:2007jg} it has been shown that the evaporation ends with non-zero value of $dM/dt$ while specific heat vanishes. In this regard it may be mentioned that it is still an open problem to find the precise condition for the remnant.

    It may be worthwhile to point out that so far we have restricted our calculation upto one loop back reaction effect in the spacetime. Correspondingly the respective values for $r_H$ have been obtained for $\eta\simeq\mathcal{O}(1)$ and $\eta=0$. In principle, there must be all loop corrections and Eqs. (\ref{9}) and (\ref{10}) will have higher order terms in $\hbar$. In this case one can easily show that $\eta$ takes the following form:
\begin{equation}
     \eta = \frac{26.182}{g} \Big(1-\sum_i \frac{2\hbar^i \gamma_i}{(r_H^2)^i} \Big)~. 
\label{final}
\end{equation}
    Here $\gamma_i$s are the combination of $\alpha$, $\beta$ and other dimensionless parameters appearing in the expressions for temperature and horizon radius. $i=1$ term gives the 1st order correction as mentioned in Eq.(\ref{eta1}). So including higher order terms does not change the physical nature of $\eta$. Here also one can, in principle, obtain the values of $r_H$ by solving equations $\frac{26.182}{g} \Big(1-\sum_i \frac{2\hbar^i \gamma_i}{(r_H^2)^i} \Big)= \mathcal{O}(1)$ and $\Big(1-\sum_i \frac{2\hbar^i \gamma_i}{(r_H^2)^i} \Big)=0$ corresponding to $\eta\simeq\mathcal{O}(1)$ and $\eta=0$, respectively such that $\sum_i \frac{2\hbar^i \gamma_i}{(r_H^2)^i}>0$. It is very hard to solve such polynomial equations and hence in our main discussion we restricted our calculation within the one loop. So it shows that even for all loop calculation, the physical nature of the spectrum will not change; only the numerical values that we have calculated for $r_H$ for different situations, will change.

    Finally, let us make a comment on the possible connection between the generalized uncertainty principle (GUP) and the cascade of Hawking evaporation. It has been observed that under GUP the Hawking temperature and the horizon entropy get modified \cite{Medved:2004yu,Banerjee:2010sd}. More interestingly this provides a possible way to resolve the information paradox problem since such principle does not allow the complete evaporation of the black hole; i.e. one has the remnant of black hole \cite{Banerjee:2010sd}. Therefore it has been argued that at the final stage of the black hole evaporation, the information is stored in this remnant. Hence one might be interested to see the role of the GUP in the present context. Note that in the above discussion the required parameter $\eta$ has been calculated from the distribution of the emitted particle (here it is Bose distribution, given by Eq. (\ref{n})). In presence of back reaction, since the modified Schwarzschild metric is still static, spherically symmetric, we have adopted that usual distribution as obtained for normal cases. Now the question is: Can we have the similar type of metric in the light of GUP? The answer of this is still unknown and therefore it may not be wise to use the same distribution. Also it is not known how to find out the distribution law starting solely from information of the GUP. Moreover, the information of the horizon area, which is roughly the absorption cross-section, is also needed (see Eq. (\ref{6})). Therefore it is obvious that the background metric is very important quantity in this analysis. This can be derived in the following way. Since GUP modifies the dispersion relation \cite{Banerjee:2010sd}, there is a possibility of obtaining the modified background metric by using it. The procedure will be exactly like the rainbow gravity \cite{Magueijo:2002xx}. After this one can calculate the distribution law by solving the external field equation under the modification by GUP. Then rest of the things will be followed automatically. Since it is completely a new project and as one needs to start from the scratch, we leave this for future study. But since back reaction and GUP both implies that same type of modifications in the horizon temperature and entropy, we expect that it plays the similar role in the Hawking evaporation cascade, like the back reaction. Of course to make a concrete statement, the whole analysis has to be done step by step. 

\section{\label{con}Conclusions}  
   Adopting the machinery of the recent paper \cite{Gray:2015pma}, we studied the nature of the cascade of Hawking radiation, as seen by an asymptotically infinity distance observer, for the Schwarzschild black hole in presence of back reaction in the spacetime. Depending on the value of $\gamma=\alpha+\beta$, either positive or negative or zero, it was found that three cases are possible. Among them when $\gamma\leq 0$, then there is nothing interesting; the nature of emission is similar to the situation when there is no back reaction effect. Only difference is for $\gamma<0$, the sparsity increases with the decrease of black hole mass. The completely different and new situation happens for positive values of $\gamma$. In this case we showed that the continuous cascade is possible for a particular value of mass. Also there is a mass limit from and below which $\eta\le 0$. We argued that since physically this is impossible, the black hole will stop radiating. Thereby we obtain a Planck size remnant. 

    So far we are aware of that in literature only the value of $\alpha$ is available. This is given by Eq. (\ref{alpha}). Since the precise numerical value of $\beta$ has not been obtained so far (which is beyond the scope of this paper), we can not say which bosons will satisfy $\gamma>0$ condition; which is the most interesting one among the three cases. But it is guaranteed that there is a possibility to have $\gamma>0$ at least for some bosons.   
    
    The issue have been studied here only for bosons. In principle during Hawking radiation any type of particle emission is possible. We can assure that for fermions the calculation will be same and one can easily check that the ultimate conclusion will be same. In this case only the numerical values of mass for the corresponding points (continuous cascade and no Hawking radiation points) will be different. This is because the numerical pre-factor in (\ref{eta1}) will change now. As per the nature of $\eta$ concerned, this numerical factor is not so important. It only rescales $\eta$ without affecting its nature.

   Finally, we want to mention a recent result by Hod \cite{Hod:2015wva}. He showed that even in the standard background of Kerr, a continuous cascade is possible if one considers super-radiant modes. In the present one since the spacetime is the Schwarzschild, there is no question of super radiation. Also as correctly pointed in \cite{Gray:2015pma}, as far as Hawking radiation is concerned, we do not need to take into account the super-radiant modes, because these are distinct from Hawking radiation modes. In this respect the current analysis as well as the results are completely new. We hope the study will illuminate this area of black hole emission.    
    
\vskip 9mm
\section*{Acknowledgments}
The research of one of the authors (BRM) is supported by a START-UP RESEARCH GRANT (No. SG/PHY/P/BRM/01) from Indian Institute of Technology Guwahati, India.

\end{document}